\documentclass[fleqn,usenatbib,a4paper]{mnras}

\usepackage{savesym}
\usepackage{amsmath}
\savesymbol{iint}
\usepackage{txfonts}
\restoresymbol{TXF}{iint}

\usepackage[T1]{fontenc}
\usepackage{ae,aecompl}

\usepackage{graphicx}	
\usepackage{amssymb}	
\usepackage{xspace}
\usepackage{enumitem}

\newcommand{\hst}{{\it HST}\xspace}

\title[Globular clusters in the FUV]{Globular clusters in the far-ultraviolet: evidence for He-enriched second populations in extra-galactic globular clusters? }

\author[M.B. Peacock et al.]{
Mark B. Peacock$^{1}$\thanks{E-mail: mpeacock@msu.edu (MBP)} 
Stephen E. Zepf$^{1}$ 
Arunav Kundu$^{2}$ 
and Julia Chael$^{1}$ 
\\
$^{1}$Department of Physics and Astronomy, Michigan State University, East Lansing, MI 48824, USA \\
$^{2}$Eureka Scientific, Inc., 2452 Delmer Street, Suite 100 Oakland, CA 94602, USA
}

\date{Accepted 2016 September 15}

\pubyear{2016}

\begin{document}
\label{firstpage}
\pagerange{\pageref{firstpage}--\pageref{lastpage}}
\maketitle

\begin{abstract}
\label{sec:abstract}

We investigate the integrated far-ultraviolet (FUV) emission from globular clusters. We present new FUV photometry of M~87's clusters based on archival \hst WFPC2 F170W observations. We use these data to test the reliability of published photometry based on \hst STIS FUV-MAMA observations, which are now known to suffer from significant red-leak. We generally confirm these previous FUV detections, but suggest they may be somewhat fainter. We compare the FUV emission from bright ($M_{V} < -9.0$) clusters in the Milky Way, M~31, M~81 and M~87 to each other and to the predictions from stellar populations models. Metal-rich globular clusters show a large spread in $FUV-V$, with some clusters in M~31, M~81 and M~87 being much bluer than standard predictions. This requires that some metal-rich clusters host a significant population of blue/extreme horizontal branch (HB) stars. These hot HB stars are not traditionally expected in metal-rich environments, but are a natural consequence of multiple populations in clusters -- since the enriched population is observed to be He-enhanced and will therefore produce bluer HB stars, even at high metallicity. We conclude that the observed FUV emission from metal-rich clusters in M~31, M~81 and M~87 provides evidence that He-enhanced second populations, similar to those observed directly in the Milky Way, may be a ubiquitous feature of globular clusters in the local universe. Future \hst FUV photometry is required to both confirm our interpretation of these archival data and provide constraints on He-enriched second populations of stars in extra-galactic globular clusters. 

\end{abstract}




\section{Introduction}
\label{sec:intro}

Far ultraviolet (FUV) observations of early-type galaxies and globular clusters offer a unique probe into their stellar populations. Main sequence and red giant branch stars in these old stellar populations are too faint at these wavelengths to contribute significantly. Instead, the integrated FUV emission is likely dominated by the He-core burning horizontal branch (HB) stars, as observed in the Galactic globular clusters \citep{Catelan09, Dalessandro12}. Indeed, these stars are leading contenders in explaining the `UV-upturn' observed in early-type galaxies \citep[e.g.][]{OConnell99, Brown00}. 

Observations of the Galactic globular clusters have shown complex HB star morphologies. The long proposed `first parameter' that is known to influence HB star morphology is metallicity, with metal-rich globular clusters having redder HB stars \citep{Sandage60}. However, there is significant scatter around this correlation, with similar metallicity clusters having different HB morphologies \citep[e.g. NGC~288 and NGC~362][]{Bellazzini01} and some clusters showing significant tails of hotter `extreme'-HB stars \citep[which may not follow the same metallicity correlation;][]{OConnell99}. This has lead to a variety of proposed `second parameters', many of which may influence the HB stars. These include: age; He-abundance; stellar core rotation; and globular cluster core density \citep[see e.g.][]{Fusi_Pecci97, Catelan09}. 

He-enhanced stars may be expected in globular clusters. Observations have demonstrated the ubiquity of multiple stellar populations in the Galactic globular clusters \citep[see e.g.][for a review]{Gratton12}. To explain the observed abundances of light elements, current theories of the formation of second population stars invoke H-burning at high temperatures \citep[e.g.][]{DAntona07, Bragaglia10}. Therefore, a natural prediction of these theories is that the second population stars should be significantly enriched in Helium. Observations of the Galactic globular clusters appear to confirm this (see Section \ref{sec:He-enhanced}, and the references therein). These studies have demonstrated that the combination of first population stars and He-enhanced second populations can help to explain the complex HB star morphologies observed. 

In this paper, we consider the integrated FUV properties of globular clusters. Such data only exists for the globular cluster systems of the Milky Way \citep{Dalessandro12}, M~31 \citep{Rey07}, Cen~A \citep{Rey09} and M~87 \citep{Sohn06}. We review these data in Section \ref{sec:fuv_data}. Interestingly, \citet{Sohn06} observed potential ``excess'' FUV emission from M~87's clusters. However, their STIS MAMA observations have subsequently been found to suffer from significantly more red-leak than previously thought \citep{Boffi08, Biretta16}. Therefore, in Section \ref{sec:m87}, we use WFPC2 F170W observations to confirm this previous FUV photometry. In Section \ref{sec:fsps}, we consider the influence of different hot populations on the FUV emission from clusters and (in Section \ref{sec:gc_fuv}) compare these models to the observed FUV emission from clusters in the Milky Way, M~31, M~81 and M~87. We conclude in Sections \ref{sec:He-enhanced} and \ref{sec:conclusions}, where we discuss the results in the context of He-enhanced second population stars in these globular clusters.

\section{FUV observations of globular clusters} 
\label{sec:fuv_data}

It is generally more challenging to detect and calibrate FUV emission (rather than redder emission) from old stellar populations. A particular concern is red-leak, in which photons from longer wavelengths contribute to the counts in an FUV filter+detector. For old stellar populations emission in the FUV can be $<10^{-5}$ times that at redder wavelengths. Therefore, even a small sensitivity to red photons can significantly contribute to the observed emission. 

Stochastic effects are an additional complication in analysing integrated FUV observations of globular clusters. For faint clusters, single sources can provide a significant fraction of the total FUV emission. To limit these effects we restrict our analysis to only include massive globular clusters, with $M_{V}<-9.0$. Imposing this cut also allows for a fairer comparison between the Galactic globular clusters and M~87's globular clusters, all of which are detected to this limit \citep[see fig. 13 of][]{Sohn06}. 

The difficulty in detecting FUV emission from clusters with current facilities means that only few systems have such photometry. In this paper, we utilise the FUV data that is currently available. We discuss potential uncertainties in the cluster colours due to calibration difficulties and derive robust results from considering these combined datasets. 

\subsection{{\it GALEX} observations of local galaxies}

The {\it GALEX} mission has revolutionised our understanding of the ultraviolet sky. Its microchannel plate detectors are thought to have negligible red-leak and it provides well calibrated all-sky FUV and NUV photometry \citep[e.g.][]{Morrissey07}. However, it is only sensitive enough to detect globular clusters in the closest galaxies. In this paper, we utilise published {\it GALEX} photometry for: 

\begin{itemize}[leftmargin=1pc, labelsep=*, itemsep=.5em]

\item The Milky Way's globular clusters: \citet{Dalessandro12} published integrated FUV and NUV photometry from {\it GALEX} observations. Some clusters are too bright to be included in the {\it GALEX} survey. We therefore supplement this catalog with data from the Astronomical Netherlands Satellite (ANS), Orbiting Astronomical Observatory (OAO) and International Ultraviolet Explorer (IUE; as presented by \citealt{Dorman95} and taken from \citealt{Sohn06}). Optical photometry, metallicity, distances and reddening are taken from \citet{Harris96}. 

\item M~31's globular clusters: \citet{Rey07} presented FUV and NUV photometry from {\it GALEX} observations. Following \citet{Peacock11a}, we limit the sample to old confirmed clusters \citep[class 1 in][]{Peacock10a} with low extinction (with $E(B-V)<0.16$) and deredden using the reddening values presented by \citet{Fan08}. For optical photometry we use the catalog of \citet{Peacock10a} and the transformations of \citet{Jester05}. We take the distance modulus of M~31, $(m-M)_{0}=24.4$ \citep{Vilardell10}. 

\item M~81 -- GC1: {\it GALEX} FUV photometry of this massive globular cluster in M~81 was published by \citet{Mayya13}. Optical photometry is available from \citet{Santiago-Cortes10} while its metallicity ($[Fe/H]=-0.6$) and reddening ($E(B-V)=0.0$) are taken from \citet{Mayya13}. We take the distance modulus of M~81, $(m-M)_{0}=27.8$ \citep{Freedman94}. 

\end{itemize}

\noindent All of these data are calibrated to the standard {\it GALEX} AB magnitude system. These are the magnitudes used throughout this paper. The colours of all clusters are dereddened assuming $R_{FUV}=8.2$, $R_{NUV}=9.2$, $R_{V}=3.1$ and $R_{I}=1.95$ (\citealt{Cardelli89}, as used by \citealt{Dalessandro12}). We note that there is some uncertainty in the UV extinction curve. This is minimised for our extragalactic globular clusters, since they have quite low extinction, with $E(B-V)<0.16$. To include more Galactic clusters, we relax this constraint and include clusters with $E(B-V)<0.32$. However, we note that the colours may be less reliable for clusters with higher reddening.

\subsection{\hst observations }

{\it GALEX} is not sensitive enough to detect FUV emission from most of the globular clusters in more distant galaxies. For galaxies at the distance of Virgo, \hst's advanced camera for surveys/ solar blind channel (ACS/SBC), space telescope imaging spectrograph (STIS) and wide field planetary camera 2 (WFPC2) detectors are capable of detecting FUV emission from some of their globular clusters. 

Currently, integrated \hst FUV photometry has only been published for M~87's globular clusters. This detected 50 globular clusters \citep{Sohn06}. To include this photometry in our analysis, we convert from ST to AB magnitudes using the definitions of these two systems and the pivot wavelength for these STIS observation ($\lambda=1455.0{\rm \AA}$), giving: $FUV_{AB} = FUV_{ST} + 2.779$.

We wish to include this large sample of extra-galactic globular clusters in our analysis. However, these data were obtained using a different telescope/detector and calibration than the {\it GALEX} sample. Of particular concern are ``red-leak'' effects on the FUV calibration. Since the publication of the \citet{Sohn06} data, such STIS observations have been shown to suffer from significantly more contamination from red light than initially thought. In the next section, we discuss M~87's clusters in more detail, comparing the STIS photometry to that measured from WFPC2 observations to empirically test the calibration and possible red-leak contamination. 

\section{M~87's globular clusters}
\label{sec:m87} 

\subsection{Published STIS FUV photometry and red-leak} 

The vast globular cluster system of M~87 has been the focus of many studies. In this paper, we utilise the cluster catalogs of \citet[][which provides V and I band photometry from King model fits to deep \hst ACS/WFC F606W and F814W observations]{Peng09} and \citet[][which provides NUV photometry from HST/WFC3 F275W observations]{Bellini15}. 

The high number of clusters in M~87, and their low reddening, make them prime targets for \hst UV observations. \citet{Sohn06} presented the FUV magnitudes of 50 of M~87's globular clusters based on deep \hst STIS FUV-MAMA F25SRF2 observations of three fields. Comparing the $FUV-V$ colours to the Galactic globular clusters, \citet{Sohn06} observed potential `excess' FUV emission, which may suggest He-enhanced populations \citep{Kaviraj07a}. 

As noted in Section \ref{sec:fuv_data}, the calibration of such FUV photometry is challenging. The FUV MAMA detectors on both ACS and STIS were originally thought to have essentially no sensitivity to visible light photons \citep[e.g.][and references therein]{Sohn06}. However, since the publication of the STIS FUV photometry of M 87's clusters, the ACS `solar blind channel' MAMA detector was found to be significantly more sensitive to optical photons than pre-launch testing indicated \citep[e.g.][]{Boffi08}. This was subsequently confirmed to be the case in the (similarly designed) STIS FUV-MAMA detector \citep[see section 5.3.4 of][]{Biretta16}. Based on the current estimated throughput curve, the red-leak can be significant. For example, it is estimated that only $31.2\%$ of detected photons have $\lambda < 1800 {\rm \AA}$ from a G0 star with $T_{eff}=6000{\rm K}$, while the remaining (majority) of the photons have wavelengths longer than the nominal cutoff for this setup. Globular clusters are not this red, and given the currently estimated throughput, we estimate the red-leak should of the order of a few percent for typical clusters. However, \citet{Biretta16} also caution that `There is little STIS data available for confirming the wavelength or time dependence of this extra optical throughput' -- so the throughput to redder wavelengths could be even greater than is currently estimated.

Given these calibration uncertainties, an independent test of the STIS FUV fluxes of M87's globular clusters is valuable. The available WFPC2 F170W images can provide such a test. While the WFPC2 F170W images are less sensitive, they cover a larger area than individual STIS frames, and so have more clusters per exposure. Most importantly for this purpose, the WFPC2 F170W filter is well calibrated \citep[e.g.][]{Lim09}. Therefore, in the following section, we use the archival WFPC2 F170W images to constrain the FUV fluxes of M87 GCs.

\subsection{WFPC2 FUV observations}
\label{sec:wfpc2}

 \begin{figure}
 \centering
 \includegraphics[width=88mm,angle=0]{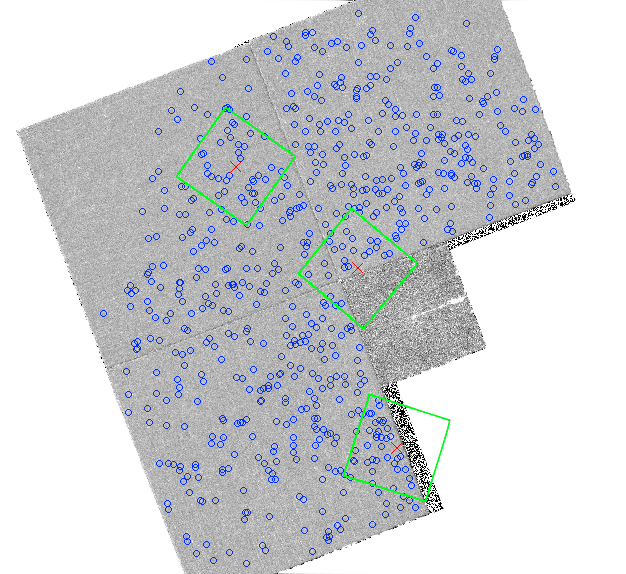} 
 \caption{WFPC2 F170W image of M~87. The green regions show the three fields observed with STIS. The blue points show the locations of clusters that are detected in the NUV and are covered by the WF chips. }
 \label{fig:fields} 
\end{figure}

M~87 was observed in 2001 using the HST/WFPC2 through the F170W filter, under proposal ID 8725. We obtained these data from the Hubble Legacy Archive\footnote{http://hla.stsci.edu} and utilise the `level 2' reduced and combined mosaic of all four of the WFPC2 detectors. Figure 1 shows this image and the locations of clusters \citep[from][blue circles]{Bellini15} and the three STIS FUV fields studied by \citet[][green polygons]{Sohn06}. It can be seen that the WFPC2 F170W field covers many of M 87's globular clusters and portions of all three STIS fields.

To compare with the published FUV, NUV and optical photometry, we need to align these WFPC2 data to a common coordinate system. We first align the bright clusters in the WFC3/UVIS F275W images to the FUV catalog of \citet{Sohn06} using the {\sc iraf} task {\sc msctpeak}. To align the F170W image to the other images we can not use the clusters, because they are very faint in the F170W observation. However, there are bright sources associated with the central jet emission \citep[discussed in detail in][]{Waters05} that are clearly detected in both F170W and F275W, and we use the central source and knots along the jet to align these images. In particular, we use these sources to determine any translational and rotation offset between the F170W image and the common coordinate system of the other datasets. Because the sources used to align the F170W images are limited to the central 17$\arcsec$ we do not allow a scaling correction, but trust in the accuracy of the WFPC2 pixel scale. The accuracy of the rotational alignment is also limited by the spatial extent of the matched sources. By stepping though different rotational offsets, we find that it is tightly constrained to be $\pm 0.3\deg$.

With only marginal detections expected in the F170W image, we performed photometry on this aligned image at the known cluster locations. We exclude clusters within 10~pixels of the chip edges and those that lie in the central PC chip, since this has significantly higher noise than the WC chips. Photometry is performed using {\sc APPHOT}, implemented under {\sc PYRAF}. We use an aperture with radius 2~pixels (0.2$\arcsec$), apply an aperture correction based on \citet{Holtzman95} and subtract a median background using an annulus with radius $4-8$~pixels. Fluxes are converted from the image units ($CTS$, electrons/s) to $F_{\nu}$ using: 
\begin{equation}
 F_{\nu} = CTS \times PHOTF_{\lambda} \times PHOTP_{\lambda} \times 10^{-18.4768}
\end{equation}
where $PHOTP_{\lambda}$ is the pivot wavelength of the filter and $PHOTF_{\lambda}$ is the flux calibration, both taken from the image headers. 

The drizzling process leads to correlated noise. We therefore utilise the inverse variance map produced by multidrizzle during the pipeline processing to estimate the noise as $\sqrt{\sum(1/{\rm IVM}_{i})}$. Here, the sum is over all pixels in the aperture and ${\rm IVM}_{i}$ is the pixel value in the inverse variance map. The resulting background noise is similar across the three detectors and consistent with that calculated from photometry at 10000 random locations on each detector, $eF_{\nu}=2.2\times 10^{-30}erg/s/cm^{2}/Hz$. This implies a $3\sigma$ detection limit of $m_{AB}=24.35$. We note that this detection limit does not account for CTE effects and the sensitivity is poorer for sources further from the readout side of the detector (those with high y pixel values; see Section \ref{sec:wfpc2_comp}). 

We also consider the effect of the uncertainty in the rotational alignment of the F170W images on our photometry. As noted above, the nucleus and knots of the M~87 jet provide good translational alignment between F170W and previous work. However, the rotational alignment from the jet is limited to $\pm0.3 \deg$. At large radii from the centre,  rotation at this level can cause significant shifts in the cluster locations and effect the photometry. To test for this, we performed photometry at locations rotated around the central jet from -0.3 deg to 0.3 deg in steps of 0.05 deg. We then determined the summed F170W flux from the 9 NUV brightest clusters at each rotational step. The resulting emission is peaked around the best alignment from the jet (0 deg), confirming that the datasets are optimally aligned at this angle. Therefore, we adopt the photometry obtained at this rotation.

\subsection{F170W flux vs. STIS predictions}
\label{sec:wfpc2_comp}

 \begin{figure}
 \centering
 \includegraphics[width=80mm,angle=270]{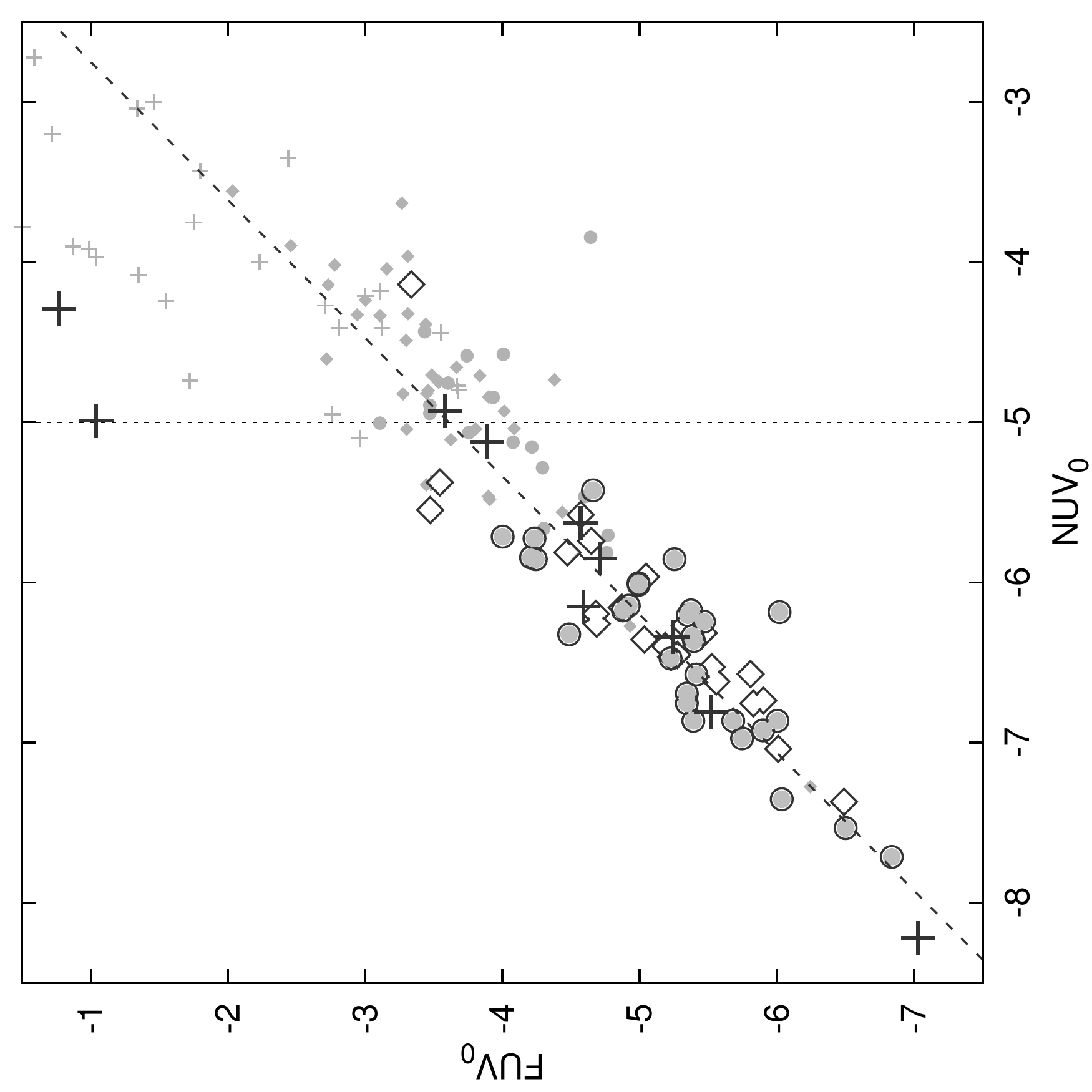} 
 \caption{$M_{FUV}$ as a function of $M_{NUV}$ for globular clusters with $M_{V}<-9.0$ in the Milky Way (plusses), M~31 (diamonds) and M~87 (grey circles). Small grey plusses, diamonds and circles show the fainter clusters in these galaxies. The dashed black line is a linear fit to all of the clusters with $NUV<-5$. }
 \label{fig:nuv_fuv} 
\end{figure}

Here, we consider how our F170W photometry compares with the published FUV photometry of \citet{Sohn06}. For clusters that are not covered by the three STIS fields, we predict their FUV flux based on their NUV fluxes and the observed relationship between FUV and NUV for M~87's clusters which have such data. This is plotted in Figure \ref{fig:nuv_fuv}. While there is significant scatter, a trend is observed, which we use to predict the FUV emission: $FUV = 1.159 \times NUV + 2.185$. For each cluster, we reduce the predicted flux to account for the charge transfer efficiency (CTE) of the WFPC2 detector. This correction is location dependent and can be significant for our WFPC2 photometry, given the very low source and background counts. For each cluster we calculate their x,y locations on the detectors by referencing to the original individual detector images. CTE corrections are then calculated using the formulae presented in \citet{Dolphin09}. The corrections vary from around 0.02~mag (at low y locations on the detector) to over 2.0~mag (at high y). 

\begin{figure}
 \centering
 \includegraphics[width=80mm,angle=0]{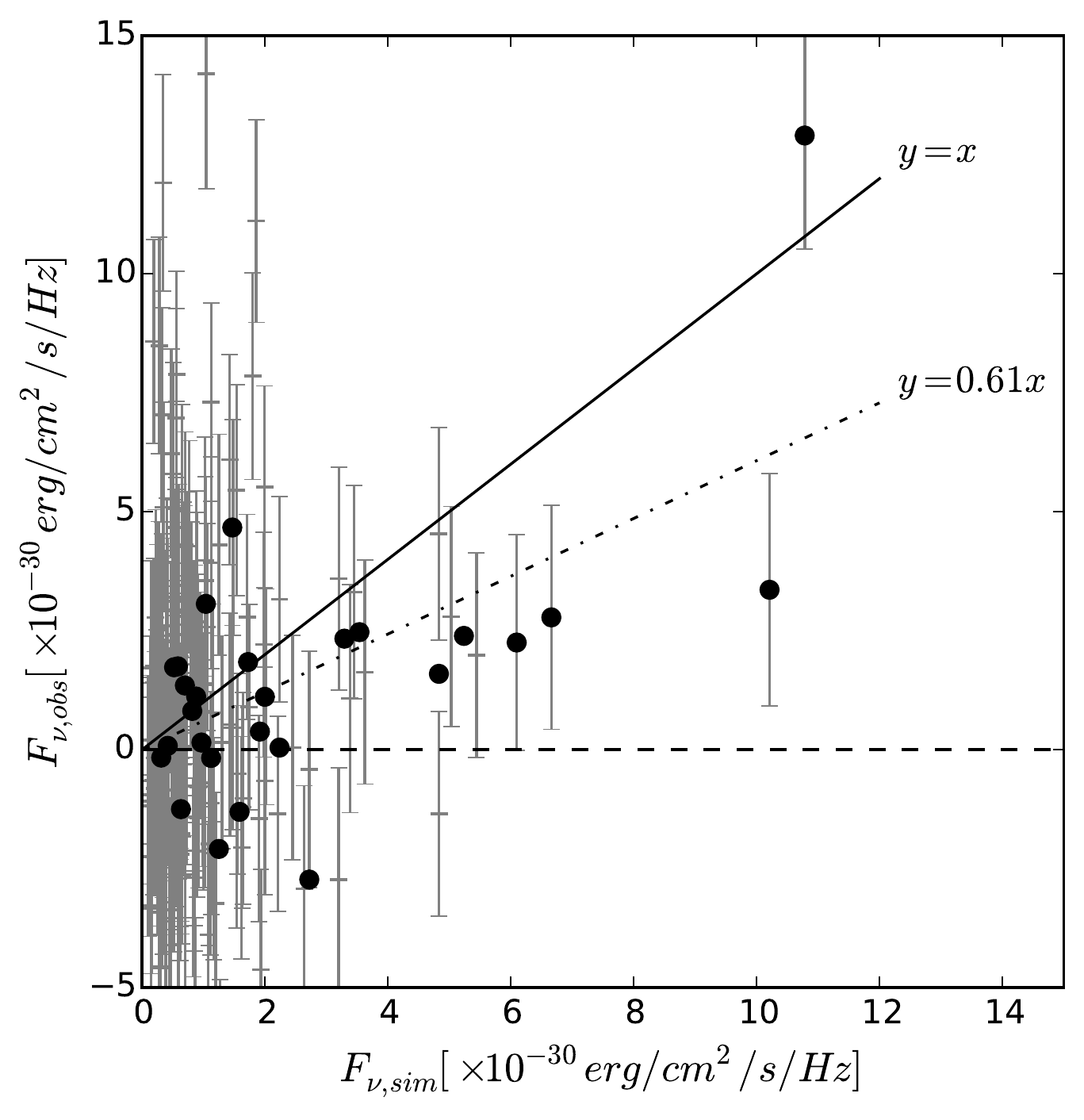} 
 \caption{Observed F170W flux ($F_{\nu, obs}$) as a function of that predicted based on observed NUV flux and the correlation presented in Section \ref{sec:wfpc2_comp}. The grey errorbars show all clusters. The black circles show the median emission from binned data. The solid $y=x$ line is the trend expected if the fluxes agree with the predictions. The dot-dashed line is the best fit to the data ($y=0.60x$). }
 \label{fig:Fnu_sim} 
\end{figure}

\begin{figure}
 \centering
 \includegraphics[width=80mm,angle=0]{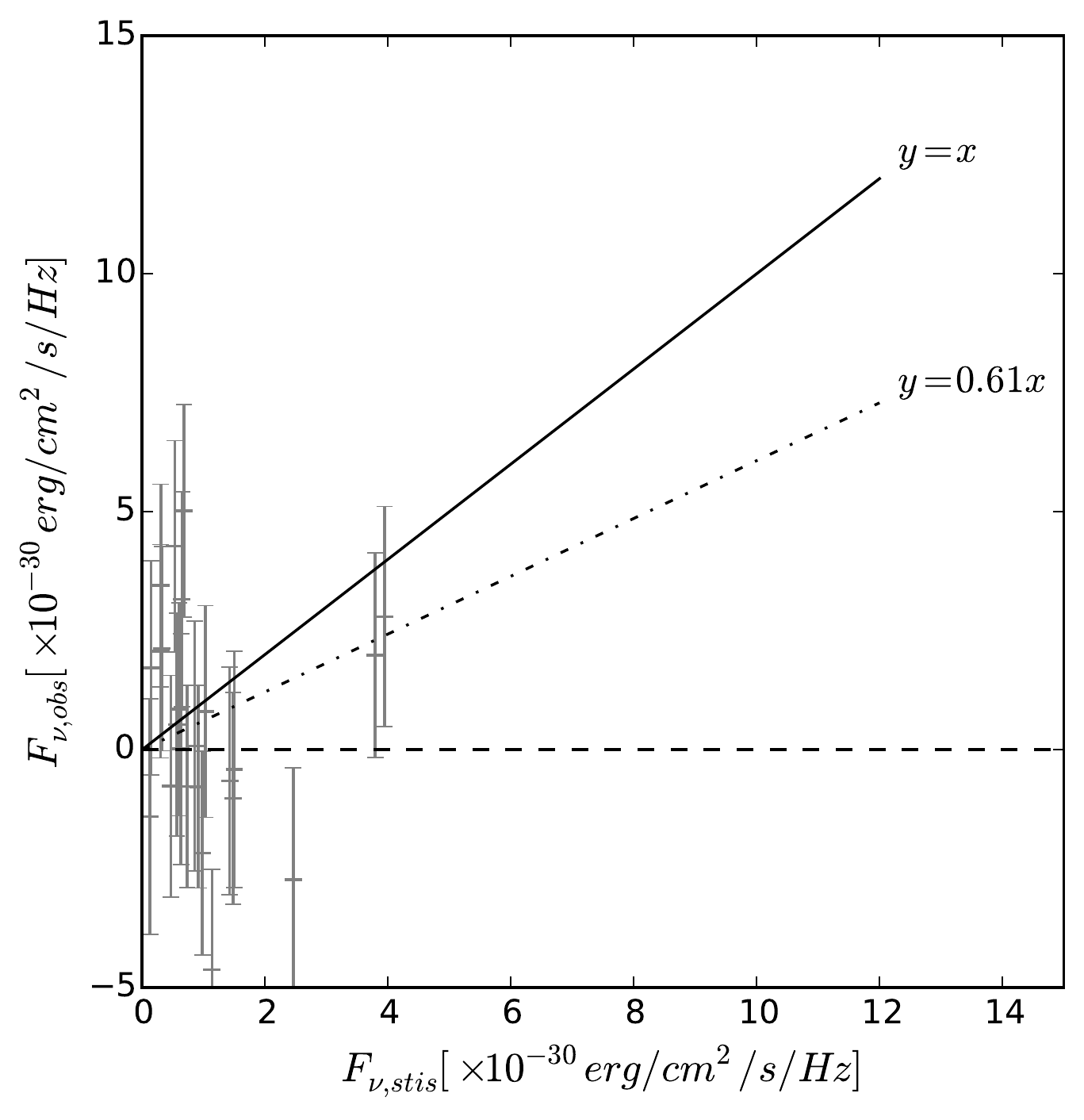} 
 \caption{Same as Figure \ref{fig:Fnu_sim}, but where the predicted flux is that directly measured from STIS images. }
 \label{fig:Fnu_stis} 
\end{figure}

In Figures \ref{fig:Fnu_sim} and \ref{fig:Fnu_stis}, we plot the measured F170W flux as a function of the predicted flux based on UVIS NUV and STIS FUV, respectively. The grey errorbars show all clusters, while the black circles show the median emission from variable bins, defined so that the sum of sources in each bin is $>6\times 10^{-30} erg/s/cm^{2}/Hz$. Ten of the brightest NUV clusters show detections at around 1$\sigma$, two of these were directly observed in the FUV with STIS. A single NUV bright source is detected at around 5$\sigma$. 

Both the NUV and FUV predictions are in reasonable agreement with the observed fluxes. However, our best fit to these data (dot-dashed line) suggests that the sources are around 60$\%$ fainter than predicted. These lower fluxes suggest a correction of about +0.5 mag should be applied to the globular clusters in the STIS FUV photometry. We note that there is some uncertainty in this correction due to the large CTE correction and the marginal detection of sources in the WFPC2 data. Future \hst FUV ACS/SBC observations using the now recommended dual filter approach are required for a more precise calibration of these magnitudes. 

\section{Stellar populations models in the FUV}
\label{sec:fsps}

To gain insights into the stellar populations of these clusters from their FUV emission, we construct models using the flexible stellar population synthesis models \citep[FSPS;][]{Conroy09, Conroy10}. We implement the FSPS code\footnote{http://github.com/cconroy20/fsps} under python-FSPS\footnote{http://dan.iel.fm/python-fsps} and use the PADOVA stellar isochrones and BASEL stellar libraries. We simulate simple stellar populations with \citet{Chabrier03} like IMFs and single ages of 13~Gyr. We use slightly shifted values for TP-AGB phase, with $delt=0.05$ and $dell=0.05$. These are found to better reproduce the observed $V-I$ -- metallicity relationship of the Milky Way's and M~31's clusters. 

In the old stellar populations of globular clusters, most stars are too cool to emit significantly in the FUV. Instead, blue-HB stars, post-AGB stars, and/or blue straggler stars likely dominate their integrated emission. Other hot populations are unlikely to significantly influence the integrated emission because they are either too faint (white dwarfs, cataclysmic variables) or too rare (X-ray binaries). Modelling of these three key populations is complex. We therefore use the flexibility of the FSPS models to vary these populations across a broad range and observe their influence on FUV emission. 

Blue straggler (BS) stars extend to hotter temperatures than the primordial main sequence. The specific frequency of BS stars, defined as $S_{BS}=N_{BS}/N_{HB}$, varies, but is typically observed to be in the range of $0.0<S_{BS}<1.0$ \citep{Piotto04}. Over this range the BS population has only a small effect on the predicted FUV emission and is insignificant when compared to variations introduced by the post-AGB and HB stars. In the models considered below, we fix $S_{BS}={\rm sbss}=0.2$.

Post-AGB stars can be extremely luminous in the FUV. However, they are short lived and therefore rare \citep[][]{deBoer87, Jacoby97, Moehler10}. Modelling this phase of stellar evolution is complex and there are uncertainties in their number, luminosity, and level of self-extinction from circumstellar dust. The FSPS code combines these uncertainties in to a weighting parameter, which allows post-AGB stars to vary from none to the full prediction based on \citet{Vassiliadis94}. Varying this weighting from 0.1 to 1.0 has only a small effect on the FUV emission from clusters with blue-HB stars, since these more numerous stars dominate the integrated emission. However, clusters with only red-HB stars are very faint in the FUV and post-AGB stars can have a noticeable influence, with $FUV-V$ getting bluer with increasing post-AGB weighting, varying from around 8 to 6 for weightings of 0.1 to 1.0, respectively. This variation is still less than that produced by the HB stars, where the FUV increases by four magnitudes from purely red to blue-HB morphologies. Lower weightings for the post-AGB stars may be more realistic, with observations of M32 showing far fewer post-AGBs than predicted \citep{Brown08} and observations of the bulge of M~31 showing that they contribute only a small fraction to the integrated light \citep{Rosenfield12}. Our final models have a weighting on the post AGB, pagb = 0.1, but we note that these stars could increase the FUV emission from red-HB dominated clusters.  

\begin{figure*}
 \centering
 \includegraphics[width=106mm,angle=270]{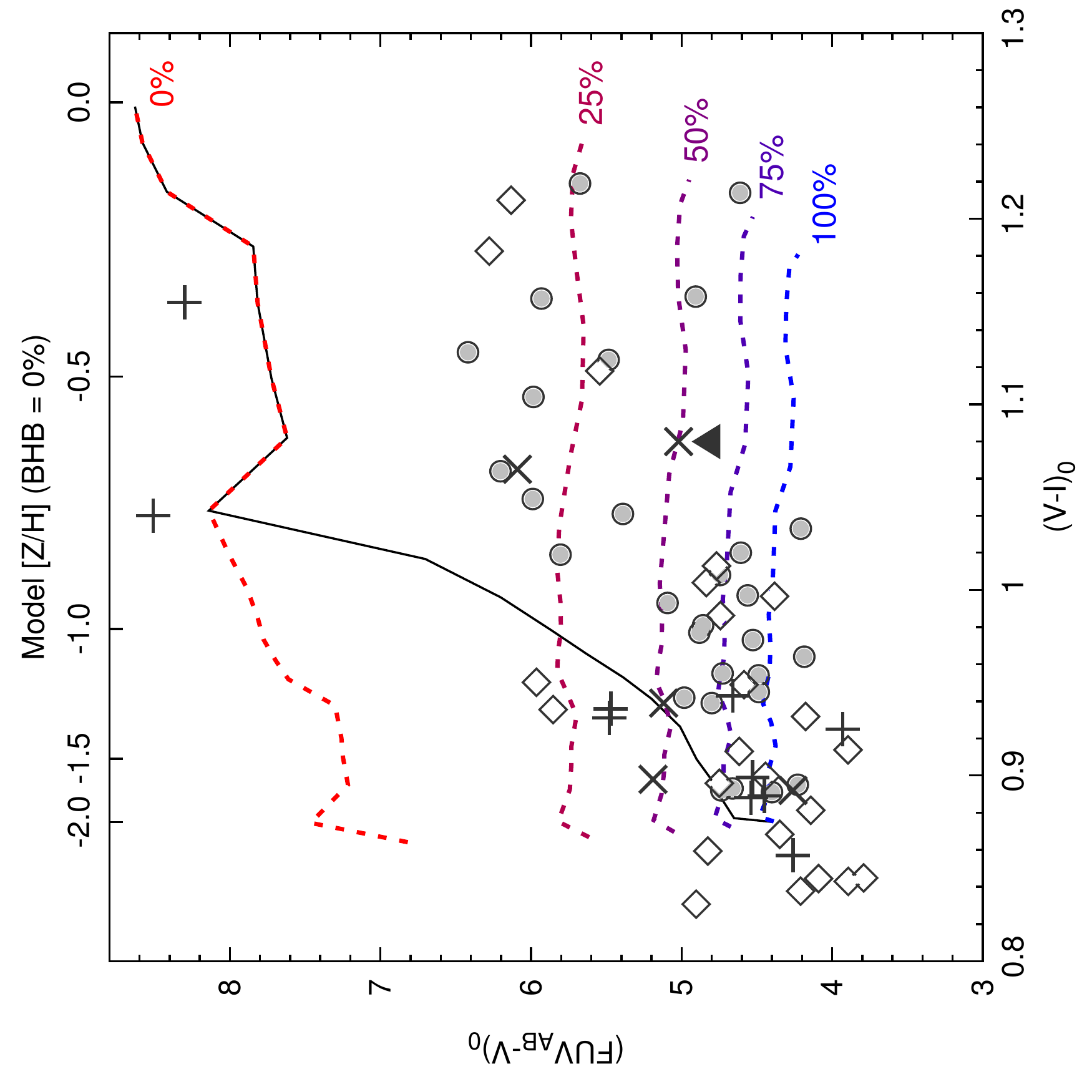} 
 \caption{$FUV-V$ as a function of $V-I$ (and corresponding metallicity for the BHB = $0\%$ model, top axis). The different symbols are for globular clusters in the Milky Way (from {\it GALEX}, plusses; and from IUE, crosses), M~31 (diamonds), M~81 (black triangle), and M~87 (grey circles). The $V-I$ colour of the Milky Way's clusters are calculated from [Fe/H] using the models in Section \ref{sec:fsps}. For M~87's clusters we shift their $FUV-V$ colour by +0.5 mags, as suggested by our comparison to the F170W photometry (see Section \ref{sec:wfpc2_comp}). The dotted lines show model predictions for clusters with blue-HB fractions of 0, 25, 50, 75, and 100\% from top (red line) to bottom (blue line), respectively. The black-solid line shows the predicted colours for clusters with a variable-HB morphology defined by Equation \ref{eq:var-bhb}. The metallicity of all of the models increases from -2.0 (left) to 0.0 (right). } 
 \label{fig:wfpc2cor} 
\end{figure*}

Blue/extreme-HB branch stars are thought to be the dominant hot populations in the old stellar populations of globular clusters and early-type galaxies. In the Galactic globular clusters, the HB star morphology is known to vary dramatically. Hot, blue-HB stars dominate some clusters while others have relatively cool, red-HB stars. In addition some clusters show a long `tail' of extreme-HB stars extending to very high temperatures. The FSPS code allows the HB star properties to be controlled via by setting the fraction of HB stars that are assigned to the blue-HB. We note that the temperature of the blue-HB stars cannot be controlled, and therefore bluer populations of extreme-HB stars can exist in real globular clusters. In Figure \ref{fig:wfpc2cor}, we show how $FUV-V$ varies as a function of metallicity for different HB morphologies. The dotted lines show the model predictions for clusters in which the fraction of blue-HB stars varies from $0\%$ to $100\%$ from top (red) to bottom (blue), respectively. In addition, the solid-black line shows a model in which the fraction of blue-HB stars increases with metallicity, such that: 

\begin{equation} \label{eq:var-bhb}
f_{BHB} = \begin{cases}
    -0.75 \times [Z/H] &, \ [Z/H]<-0.7 \\
    \ \ 0.0 &, \ [Z/H]\geq-0.7  \\
 \end{cases}
\end{equation}
This is the form presented by \citet{Conroy10} to fit integrated $NUV-V$. This transition from a blue- to red-HB with increasing metallicity is also consistent with the `first parameter' effect observed in resolved CMDs of the Galactic globular clusters. 

We note that this variable HB model predicts similar UV colours to those based on other stellar populations models \citep[e.g. the YEPS models: ][]{Lee05, Rey07, Chung11, Chung13}. In the next Section, we consider how these predictions compare to observed globular cluster colours.

\section{FUV properties of globular clusters}
\label{sec:gc_fuv}

Figure \ref{fig:wfpc2cor} shows the $FUV-V$ colour of globular clusters as a function of their $V-I$ colour (and corresponding metallicity, based on the stellar populations models presented in Section \ref{sec:fsps}). We include data for globular clusters from the Milky Way (where $V-I$ is calculated from [Fe/H], plusses and crosses), M~31 (diamonds) and GC1 in M~81 (black triangle). For details of these cluster data, see Section \ref{sec:fuv_data}. We also plot M~87's globular clusters (grey circles). These are based on the published STIS FUV photometry, converted to AB mags, and adjusted assuming the correction suggested in Section \ref{sec:wfpc2_comp} ($FUV=FUV_{STIS}+0.5$). The dotted lines show the model predictions for clusters with different HB morphologies (see Section \ref{sec:fsps} for details). 

We limit our analysis to bright globular clusters, with $M_{V}<-9.0$. All of M~87's globular clusters are detected to this limit, making for a fairer comparison between the Milky Way and M~87. Importantly, imposing this cut also reduces stochastic effects from single bright FUV sources by ensuring a relatively large stellar population. 

\subsection{Metal-poor clusters} 

The metal-poor globular clusters have quite similar colours, with $FUV-V\sim4.5$, and are well described by the simple stellar populations model with a blue-HB morphology. This is consistent with resolved CMDs of the Galactic globular clusters, which show that metal-poor clusters have well populated blue-HBs. We note that the models only go down to metallicities of $[Z/H]=-2.0$ and that some clusters have lower metallicities and hence extend to bluer colours. 

Comparing cluster systems suggests that M~87's clusters have similar $FUV-V$ colours, if we apply the proposed correction based on our F170W images. We therefore propose that red-leak (or other calibration issues) in the STIS photometry {\it may} account for some of the `excess FUV' emission previously noted from M~87's clusters \citep{Sohn06, Kaviraj07a}. 

\subsection{Metal-rich clusters: some are FUV bright}

The intermediate/ metal-rich globular clusters show a broad spread in $FUV-V$, with each galaxy hosting metal-rich clusters that are unexpectedly bright in the FUV. The FUV brightness of M~87's metal-rich clusters can not be explained by red-leaks, since some are as bright as their metal-poor counterparts and they show a broad spread in $FUV-V$. Additionally, we note that there are three metal-rich clusters in M~31 that are relatively FUV bright and GC1, the brightest (and metal-rich) cluster in M~81, is also FUV bright. The photometry of these four clusters is based on {\it GALEX} observations and should therefore be directly comparable to the Milky Way's clusters. 

Two metal-rich Galactic globular clusters, 47~Tuc and NGC~6356, are very faint in the FUV with $FUV-V \sim 8.0$. This is consistent with the traditional HB correlation, where metal-rich clusters have primarily red-HB stars. However, the Milky Way also contains the metal-rich clusters NGC~6388 ([Fe/H]=-0.6) and NGC~6441 ([Fe/H]=-0.53) which have $10-20\%$ blue-HB stars \citep[e.g.][]{Busso07}. Unfortunately, these clusters do not have {\it GALEX} photometry from \citet{Dalessandro12}. However, they are plotted in Figure \ref{fig:wfpc2cor} based on older IUE observations. These data suggest that their $FUV-V$ colours are approximately 2 and 3~mag bluer than 47~Tuc. 

As noted by \citet{Sohn06}, it is interesting that no 47~Tuc analogs (very FUV faint metal-rich clusters) are observed. M~31's clusters are not complete to our magnitude cut of $M_{V}<-9.0$, so similarly FUV faint clusters may be undetected. However, M~87's metal-rich clusters are mostly complete to this limit and have $FUV-V<6.5$. It is possible that red-leak may have a greater effect on the $FUV-V$ colour of truly FUV faint clusters (since less real light is emitted in the FUV). We are unable to investigate this with the data available (since our WFPC2 photometry detects only the brightest UV sources). Future observations could conclusively test for FUV faint clusters, and constrain red leaks, by using \hst with the now recommended dual FUV filter approach. 

Considering the larger sample of metal-rich clusters that are available by including extra-galactic globular cluster systems, we conclude that metal-rich clusters are not uniformly faint in the FUV, but rather span a broad range of $FUV-V$ colours. Despite considering other hot stellar populations in these clusters, the models discussed in Section \ref{sec:fsps} can only produce the FUV brightest metal-rich clusters if they host a significant population of blue-HB stars ($\sim50\%$ for these models).

\section{Helium enhanced second populations in globular clusters}
\label{sec:He-enhanced}

The broad spread in $FUV-V$ of the metal-rich globular clusters can only be explained by these stellar populations models if their HB morphology varies. The reason for this variation may be Helium enhancement. A Helium enhanced stellar population will have bluer HB stars and extend to the extreme-HB for higher He-abundances \citep[e.g.][]{Lee05}. This effect dominates over metallicity and can therefore produce blue/extreme-HB stars in metal-rich environments. 

There is compelling evidence that the Galactic globular clusters host multiple populations of stars, hosting both primordial and enriched populations \citep[see e.g.][for a review]{Gratton12}. Theories to explain the observed abundances of second population stars generally require that they form from gas enriched by H-burning at high temperatures. Therefore, a key expectation is that the second population stars are significantly enriched in He \citep[e.g.][]{DAntona02, DAntona04, DAntona07, Bragaglia10}. Recent studies suggest that such HB (as well as giant, sub-giant and main sequence) stars are He-enriched. This is based on both spectroscopy of a small number of cluster stars \citep[e.g.][]{Pasquini11, Marino14} and on fitting resolved \hst CMDs \citep[e.g.][]{Milone12b, Marino14, Piotto13, Milone15}. While current samples are small, the fraction of second population stars, and their He-enrichment, appears to vary significantly in different clusters (with $0.01<\Delta Y<0.12$). Additionally, this enrichment may correlate with cluster mass \citep[][and references therein]{Milone15}. 

The spread in the $FUV-V$ of the metal-rich clusters in Figure \ref{fig:wfpc2cor} is therefore expected if these clusters host multiple populations (like the Galactic globular clusters). The relatively FUV faint clusters are consistent with being dominated by red-HB stars due to their high metallicity {\it and} low He-enhancement \citep[like 47~Tuc, whose second population has one of the lowest He-enhancements,][]{Milone12}. While the bluest $FUV-V$ metal-rich clusters can result from them hosting significant fractions (around $50\%$) of He-enriched second populations, which produce very blue/extreme-HB stars. We note that, for primordial Helium abundance, metal-poor populations already produced quite blue-HB stars. Therefore, the effects of a Helium enhanced population on the integrated FUV emission are likely to be more pronounced in the metal-rich clusters. 

We conclude that significant fractions of blue-HB stars (a natural result of a Helium enhanced second population) are required to explain the FUV emission from some metal-rich globular clusters. This fraction is found to vary up to $\sim50\%$. We caution that the exact fraction of second population stars is dependent on the details of the stellar populations models and the temperature distribution of the blue-HB stars. We also note that the fraction of Helium enhanced second population stars in these clusters may be higher than the fraction of blue-HB stars, due to details in the evolution to the HB for different Helium abundances \citep[see e.g.][]{Caloi07, DAntona08}.

\section{Conclusions}
\label{sec:conclusions}

\begin{itemize}[leftmargin=1pc, labelsep=*, itemsep=.5em]

\item We analyse archival WFPC2 F170W far-ultraviolet (FUV) observations of M87's globular clusters. These data confirm the general FUV detection of M87's clusters from earlier work with the STIS FUV-MAMA detector and F25SRF2 filter. 

\item Our analysis of these F170W data suggests that the STIS FUV magnitudes may be $\sim 0.5$ mag too bright. This might be due to calibration uncertainties, such as a red-leak in the STIS observations, which was discovered only after publication of the initial STIS results. This shift produces $FUV-V$ colours of metal-poor clusters that are similar to those of the Milky Way, M~31 and models. However, we caution on the accuracy of this correction based on these marginal WFPC2 detections and propose that future \hst observations using a dual FUV filter approach can more accurately measure the FUV magnitudes of these and other extra-galactic globular clusters. 

\item Metal-rich globular clusters show a broad spread in $FUV - V$. This includes clusters whose $FUV - V$ colours are much bluer (by $2-4$ mags) than the predictions from standard models. We note that metal-rich clusters with blue $FUV-V$ colours are found in all galaxies we consider (M~87, M~31, and M~81) and based on data from different telescopes and instruments ({\it GALEX} and \hst). 

\item The blue $FUV-V$ colours of some metal-rich globular clusters require that they host blue/extreme-HB stars, unlike the classical expectation for a simple stellar population, where metal-rich globular clusters are expected to have a red-HB star morphology. 

\item We propose that He-enhanced second population stars are the natural source for these hot HB stars in metal-rich globular clusters -- since He-enhanced populations produce bluer HB stars, even at high metallicity. 

\item FUV observations of metal-rich clusters therefore have the power not only to test for second populations using the integrated light from globular clusters, but also to constrain He-enhancement and factors that drive its variation. 

\end{itemize}

\section*{Acknowledgements}

We thank the anonymous referee for taking the time to provide informed and helpful comments on this manuscript. We also thank Andrea Bellini for providing a copy of their M~87 cluster catalog and UVIS images and Nate Bastian for helpful discussions. Support for this work was provided by NASA through the ADAP grant number NNX15AI71G and by the NSF through grant number NSF AST-1412774 . JC's research was supported by the MSU HSHSP program. 

Based on observations made with the NASA/ESA Hubble Space Telescope, and obtained from the Hubble Legacy Archive, which is a collaboration between the Space Telescope Science Institute (STScI/NASA), the Space Telescope European Coordinating Facility (STECF/ESA) and the Canadian Astronomy Data Centre (CADC/NRC/CSA). This research has made use of NASA's Astrophysics Data System.




\bibliographystyle{mnras}
\bibliography{bibliography_etal}



\bsp	
\label{lastpage}
\end{document}